# REVISITING THE THERMODYNAMICS OF HARDENING PLASTICITY FOR UNSATURATED SOILS

| | |
|---|---|
| Olivier Coussy | Professor. Université Paris-Est, UR Navier, École des Ponts ParisTech, Marne-la-Vallée, France. |
| Jean-Michel Pereira | Researcher. Université Paris-Est, UR Navier, École des Ponts ParisTech, Marne-la-Vallée, France. |
| Jean Vaunat | Professor. Department of Geotechnical Engineering and Geosciences, Universitat Politècnica de Catalunya, Barcelona, Spain. |

**Corresponding author:**
Olivier Coussy
Université Paris-Est
UR Navier
Ecole Nationale des Ponts et Chaussées
6-8 avenue Blaise Pascal
77455 Marne-la-Vallée cedex 2
France
Email: olivier.coussy@enpc.fr
Phone: + 33 1 64 15 36 22
Fax: + 33 1 64 15 37 41

# REVISITING THE THERMODYNAMICS OF HARDENING PLASTICITY FOR UNSATURATED SOILS

ABSTRACT: A thermodynamically consistent extension of the constitutive equations of saturated soils to unsaturated conditions is often worked out through the use of a unique 'effective' interstitial pressure, accounting equivalently for the pressures of the saturating fluids acting separately on the internal solid walls of the pore network. The natural candidate for this effective interstitial pressure is the space averaged interstitial pressure. In contrast experimental observations have revealed that, at least, a pair of stress state variables was needed for a suitable framework to describe stress–strain–strength behaviour of unsaturated soils. The thermodynamics analysis presented here shows that the most general approach to the behaviour of unsaturated soils actually requires three stress state variables: the suction, which is required to describe the invasion of the soil by the liquid water phase through the retention curve; two effective stresses, which are required to describe the soil deformation at water saturation held constant. However a simple assumption related to the plastic flow rule leads to the final need of only a Bishop-like effective stress to formulate the stress–strain constitutive equation describing the soil deformation, while the retention properties still involve the suction and possibly the deformation. Commonly accepted models for unsaturated soils, that is the Barcelona Basic Model and any approach based on the use of an effective averaged interstitial pressure, appear as special extreme cases of the thermodynamic formulation proposed here.

## INTRODUCTION

One of the early applications of the mathematical theory of plasticity to soil mechanics goes back to the pioneering work of Roscoe and his co-workers on the general concept of critical state (Roscoe *et al.*, 1958; Roscoe *et al.*, 1963), which, for saturated soils, ultimately resulted in the elaboration of the celebrated Cam-Clay model involving Terzaghi's effective stress. An extension of the Cam-Clay model to unsaturated conditions has been further proposed by Alonso *et al.* (1990). Within a simple elastoplastic formalism this extension has pointed out the need of two stress state variables instead of a unique effective stress in order to account for experimental observations on the mechanical behaviour of unsaturated soils. As a consequence, this model has launched the bases of many models further developed for unsaturated soils and addressing additional aspects, such as the effects of the Lode angle (Sun *et al.*, 2000), of water content (Wheeler, 1996; Vaunat *et al.*, 2000), of anisotropy (Cui & Delage, 1996; Ghorbel & Leroueil, 2006) and of the degree of saturation (Jommi *et al.*, 1994; Bolzon *et al.*, 1996; Dangla *et al.*, 1997; Lewis and Schrefler, 1998; Gallipoli *et al.*, 2003; Sheng *et al.*, 2004; Pereira *et al.*, 2005; Sun *et al.*, 2007; Wheeler *et al.*, 2003). In turn advances performed on the last point have reintroduced a strong debate about the possible relevancy of the effective stress concept to capture the mechanical behaviour of unsaturated soils, an issue dated back to the 1960s.

The concept of effective stress for unsaturated soils takes its roots in the work by Bishop (1959) that extended the concept of Terzaghi's effective stress by replacing the water pressure by the weighted average of air and water pressures according to:

$$\sigma_{ij}^{B} = \sigma_{ij} - [u_a + \chi (u_a - u_w)] \delta_{ij}$$

where $\sigma_{ij}^{B}$ and $\sigma_{ij}$ are the Bishop's and total stress tensors, respectively, while $u_a$ is the air pressure, $u_w$ the water pressure and $\chi$ a weighting parameter. The dependence of this weighting parameter $\chi$ upon the degree of saturation of water $S_r$ has been stated in the early work of Bishop and co-workers (Bishop, 1959; Bishop & Blight, 1963). Historically, no definitive statement about how this parameter depends on $S_r$ has been made for more than 20 years after Bishop's proposal. Lewis & Schrefler (1982) followed by Bear *et al.* (1984), cited by Bear & Bachmat (1990), used the natural candidate $\chi(S_r) = S_r$ as a special case for the weighting factor (deduced from volume average of the pressures of the fluids saturating the porous space). Some other proposals have been made. For instance, Khalili & Khabbaz (1998) identified $\chi$ as a function of the suction from experiments performed on the shear strength of a large set of soils. This work has been used later on by Loret & Khalili (2002).

Even though the choice of $\chi(S_r) = S_r$ is natural and largely used, the status of the weighting function $\chi$, as well as the choice of its relevant argument, remain unclear. The formulation of the constitutive equations of saturated soils using Terzaghi's effective stress relies on the incompressibility of the solid grains. Hereafter, this grain incompressibility being a starting point, it will be shown how thermodynamics can bring answers to the question of using an effective stress regarding the constitutive equations of unsaturated soils. It will be in particular revealed that significant assumptions related to the flow rule are actually needed to validate the natural choice $\chi(S_r) = S_r$.

Another attractive approach to explore the concept of effective stress for unsaturated porous media is provided by averaging methods. For elastic porous solids, using homogenization techniques, Chateau & Dormieux (2002) have shown the relevancy of adoption $\chi(S_r) = S_r$ if the strain localization tensor is the same for all pores, which turns out to assume the iso-deformation of all pores. Without having recourse to these sophisticated methods, the consequences of this iso-deformation assumption will be revisited in this paper in the context of plasticity directly at the macroscopic scale. An alternative to homogenization procedures is the so-called hybrid-mixture theory. The latter establishes balance equations at the microscopic scale and performs a change of scale through averaging techniques (Hassanizadeh & Gray, 1980; Murad *et al.*, 1995; Lewis & Schrefler, 1998). Such averaging methods offer the advantage to provide a direct interpretation of the macroscopic variables in terms of their microscopic counterparts. Particularly, if volume average is considered, an equivalent interstitial pressure equal to the product of the degree of saturation by suction comes out from the analysis. However the constitutive laws are then usually developed at the averaged scale from considerations based upon the entropy inequality which, as an inequality, cannot offer a definitive answer. Recently, Gray & Schrefler (2007) have replaced in a thermodynamic context the use of Bishop's stress in its original form in identifying the parameter $\chi$ to the fraction of the solid phase surface area in contact with the wetting phase.

The actual question is to determine whenever the choice $\chi(S_r) = S_r$ is relevant. As recalled above micromechanics shows that the choice is relevant providing all of the pores undergo the same dilation whenever the same pressures apply to their solid walls. For elasticity and plasticity, restricting to macromechanics this paper will explore as far this choice is relevant. Regarding observations, the experimental evidence of the concept of an effective stress for unsaturated soils has often been questioned. One among the most employed arguments on the limitations of this concept is that it cannot account for the collapse occurring along wetting paths under constant stress (e.g. Jennings & Burland, 1962; Matyas & Radakrishna, 1968). This well-known phenomenon is characterised by plastic compression, possibly preceded by elastic swelling during the soaking of an unsaturated soil sample under constant stress. This recurrent criticism to the various expressions of the effective stress appearing from time to time in the literature lies on the fact that such a stress cannot reproduce alone the response of the material and, thus, departs from the historical definition given by Terzaghi (1936): "*All the measurable effects of a change of stress, such as compression, distortion and a change in the shearing resistance are exclusively due to changes in effective stress… every investigation of the stability of a saturated body of earth requires the knowledge of both the total and the neutral stresses*". However, as indicated by Jommi (2000), such a condition has been never met, even for saturated materials. Gens (1995) referred to the more adequate definition that any change in total stress and neutral pressures that causes the same change in effective stress traduces into the same response of the material.

In most of all the mentioned previous approaches, the key variable controlling the behaviour of unsaturated soils is the suction. Its variations directly control the fluid invasion process through the water retention curve, which is eventually associated with the surface energy balance. The suction variations also control indirectly the mechanical behaviour through the variation of the strength and of the locked energy they induce. In the familiar capillary case, although the suction can be defined as the difference between the pressures of the non-wetting and wetting phases, the various roles of the pressure difference must be well separated from that of the suction. For instance, in the case of non-connected fluid phases occupying always the same part of the porous volume, there is no invasion process by a non-wetting fluid so that the suction has no meaning, whereas the difference between the pressures of the fluids still governs the deformation of the material, with an appropriate choice of the stress variables. In the case of connected phases, this specific mechanical role of the pressures difference still remains, irrespective of that of suction previously defined which, in turn, will also affect the mechanical behaviour, as for instance the strength. A parallel can be made here between the role played by the suction upon the mechanical behaviour and the analogous role of chemical effects appearing in some reactive porous media (Coussy & Ulm, 1996). Indeed, chemo-mechanical couplings can induce variations of strength due to the chemical reactions taking place within the material, resulting in a chemical hardening similar to the capillary hardening we just evoked. Same comments apply to the influence of chemical reactions, similar to that of suction variations, upon the hardening locked energy.

At the light of these various roles played by the pressure difference, in this paper we will revisit most of the points previously described in the modelling of unsaturated soils behaviour within an elastoplastic framework. Controlling variables are first looked for by analyzing the strain work related to unsaturated materials. By means of the analysis of the strain work, a special care is devoted to the identification of the different physical processes governing the deformation and the pore invasion by fluids. The concept of effective stress is then derived through adequate dependencies in thermodynamics potentials. Finally, an illustration of the framework is presented, where the Barcelona Basic Model (BBM) for unsaturated soils is analysed and found to be thermodynamically consistent. It comes out from this analysis that any approach based on the use of an averaged interstitial pressure are eventually two special cases of a more general thermodynamic approach recently proposed (Coussy, 2007) for the formulation of the constitutive equations of unsaturated soils.

An unsaturated soil is constituted of a solid skeleton formed of particles in contact, through interfaces having their own energy, with a gas phase and a liquid phase. The thermodynamics of this solid skeleton can be addressed by considering successively three systems. The first system is the soil itself, as just depicted. This is an open system exchanging gas and liquid mass with the surroundings. The second system is obtained by removing from the soil system the bulk gas and liquid phases, whose thermodynamics is separately known. Since this system does not include any more the fluid phases, this system is formed of only the solid particles and the interfaces. That the bulk fluid phases have been removed does not mean that the system is no more subjected to the pore pressures. This system is actually a closed system which is loaded by the total external stress and the pore pressures still exerting through the interfaces on the system. In the following this system is called the apparent solid skeleton. Indeed, this is an apparent solid skeleton since the interfaces have their own energy. As a result they have also to be removed to define the actual solid skeleton whose constitutive equations are those we are looking for. We call it the solid skeleton in the following. In short three systems are considered: the soil (solid skeleton + interfaces + bulk fluid phases), the apparent solid skeleton (solid skeleton + interfaces), and finally the solid skeleton.

STRAIN WORK WITH NON CONNECTED FLUID PHASES

*Strain work in soils with one saturating fluid*

In a first instance, the case of saturated soils is briefly revisited. We consider the case of a soil under isotropic loading conditions. In the reference configuration, the material is free of any total stress. Its volume is $V_0$ and porosity $n_0$. In this configuration, pores are saturated by the liquid at zero (atmospheric) pressure. At time $t$, once applied an isotropic loading, the material in the current deformed configuration is characterised by volume $V$, mean total stress $p$, porosity $n$ and pressure of the saturating fluid $u$. Since porosity $n$ is defined relatively to the current volume $V$ as generally done in soil mechanics, it can be coined as the (usual) Eulerian porosity. By opposition we can refer the current porous volume $nV$ to the initial volume $V_0$ by writing:

$$nV = \phi V_0 \qquad (1)$$

The porosity $\phi$ can be coined as the Lagrangian porosity (Coussy, 2004) since it is defined relatively to the initial volume $V_0$. In the reference configuration, $n_0$ is equal to $\phi_0$. From time $t$ to time $t + dt$, the infinitesimal work $dW$ supplied to the solid skeleton has two contributions: the infinitesimal work of the total stress, that is $-pdV$, and the infinitesimal work of the pressure exerted by the saturating fluid on the solid walls of the porous network, that is $ud(nV)$, resulting in:

$$dW = -pdV + ud(nV) \qquad (2)$$

We now refer the infinitesimal work $dW$ to the initial volume by writing:

$$dW = V_0 \, dw \qquad (3)$$

where $dw$ is the infinitesimal strain work related to the solid skeleton. Use of (1) and substitution of (3) in (2) provides the equation for the infinitesimal strain work $dw$:

$$dw = p \, d\varepsilon_v + u \, d\phi \qquad (4)$$

where $\varepsilon_v$ is the volumetric strain:

$$\varepsilon_v = -\frac{V - V_0}{V_0} \qquad (5)$$

It should be pointed out that the Lagrangian and Eulerian porosities, respectively $\phi$ and $n$ must be distinguished. As an illustration, consider the work produced by the pore pressure. It is equal to $ud\phi$ which is equal to $u(d\phi - \phi_0 d\varepsilon_v)$ at the first order. Assuming $ud\phi = udn$ would lead to neglect a term $u\phi_0 d\varepsilon_v$ having the same order of magnitude as $pd\varepsilon_v$.

When the solid grains are incompressible, the volumetric strain $\varepsilon_v$ is only due to the changes in porous volume. Accordingly:
$$d\varepsilon_v = -d\phi \tag{6}$$
and (4) can be rewritten in the form:
$$dw = p'd\varepsilon_v \tag{7}$$
where:
$$p' = p - u \tag{8}$$
is the mean component of the Terzaghi's effective stress. Equation (7) can be extended to triaxial stress conditions by adding to the strain work the contribution associated to the deviatoric stress $q$, work conjugate variable of the deviatoric strain $\varepsilon_q$:
$$dw = p'd\varepsilon_v + qd\varepsilon_q \tag{9}$$
Equation (9) is comparable to the equation early established by Schofield & Wroth (1968).

*Strain work in soils with two non-connected saturating fluids*

Consider now the case where the porous volume is formed by two disconnected porous networks. This "non-connected" case may be unusual for soils (not for rocks). However it is introduced as an illustration of the role of the suction without yet considering its effects on the drying/wetting process. As it will be seen later, this illustrative case thus permits to introduce in a natural manner the new concept of Lagrangian saturation.

In the reference configuration (Fig. 1-left), the material is free of any total stress and interstitial pressure. Its initial volume and overall porosity are $V_0$ and $n_0 = \phi_0$ respectively. In the current configuration (Fig. 1-right), the volume is $V$, the overall Lagrangian porosity $\phi$, the mean total stress $p$ and the deviatoric stress $q$. As sketched in Fig. 1-right, one porous network is filled by air (index a) and the other by water (index w). In this configuration, the volumes occupied by the water and air phases are given by, respectively:
$$\phi_w V_0 = s_r \phi V_0 \tag{10}$$
$$\phi_a V_0 = (1 - s_r) \phi V_0 \tag{11}$$
And thus:
$$\phi_a + \phi_w = \phi \tag{12}$$
$\phi_a$ and $\phi_w$ can be coined as the partial Lagrangian porosities since they relate the current volume of air and water to the initial volume $V_0$. In soil mechanics, $\phi_a$ and $\phi_w$ may be identified to air and water volumetric contents. $s_r$ represents the fraction of the current porous volume occupied by water. Actually, since $s_r$ applies to the current porous volume $\phi V_0 = n V$, $s_r$ can be coined as the Eulerian water saturation.

Following the same reasoning as for Equation (4), the infinitesimal strain work $dw$ related to the apparent solid skeleton during time $dt$ can be expressed in the form:
$$dw = pd\varepsilon_v + q\,d\varepsilon_q + u_a\,d\phi_a + u_w\,d\phi_w \tag{13}$$
Substitution of the expressions for Lagrangian air and water volumetric contents (10) and (11) into (13) and use of Equation (12) provide the alternative expression:
$$dw = p\,d\varepsilon_v + q\,d\varepsilon_q + [(1-s_r)u_a + s_r u_w]d\phi - \phi(u_a - u_w)\,ds_r \tag{14}$$
Equation (14) is similar to the expression derived by Dangla *et al.* (1997) and Coussy (2004). In the case of a solid skeleton formed by incompressible grains, it can be rewritten by substituting (6) into (14):
$$dw = [p - (1-s_r)u_a - s_r u_w]d\varepsilon_v + q\,d\varepsilon_q - \phi(u_a - u_w)\,ds_r \tag{15}$$
which agrees with the expression obtained by Houlsby (1997). According to Equation (15), for non-connected fluid phases, the stress couple formed by Bishop's mean stress (with $\chi$ factor identified to the Eulerian water saturation $s_r$), that is $p - (1-s_r)u_a - s_r u_w$, and the pressure difference $-\phi(u_a - u_w)$ (opposite to the product matric suction $u_a - u_w$ by Lagrangian porosity $\phi$) is work conjugate to the strain couple formed by the volumetric strain $\varepsilon_v$ and the Eulerian water saturation $s_r$.

Alternative sets of work conjugate stress and strain can be identified by using Equations (12) and (13) together with the condition for solid incompressibility (6). A first possibility is to express the infinitesimal strain work as:

$$dw = (p - u_a)\, d\varepsilon_v + q\, d\varepsilon_q - (u_a - u_w)\, d\phi_w \qquad (16)$$

Equation (16) indicates that the couple formed by the net stress $p - u_a$ and the opposite of matric suction $-(u_a - u_w)$ is work conjugate to the couple formed by $\varepsilon_v$ and $\phi_w = s_r\, \phi$. An alternative option is to preserve the symmetry of the formulation with regard to both fluid phases. Bearing in mind that $d\varepsilon_v = -d\phi_w - d\phi_a$ holds when the solid skeleton is formed of incompressible grains (see Equations (6) and (12)), $dw$ can be rewritten in the form:

$$dw = (p - u_a)\, d\phi_a - (p - u_w)\, d\phi_w + q\, d\varepsilon_q \qquad (17)$$

By opposition to equations (15) and (16), equation (17) separates the contribution of the air phase from the contribution of the water phase in the expression of the infinitesimal strain work $dw$.

Equations (16) and (17) should be compared to the work of Fredlund & Morgenstern (1977) who showed that any couple of variables among $(p - u_a)$, $(p - u_w)$ and $(u_a - u_w)$ may be used as stress states variables for modelling of unsaturated soils behaviour.

STRAIN WORK WITH CONNECTED FLUID PHASES: ACCOUNTING FOR THE RETENTION CURVE

In the previous case, the strain work $dw$ was accounting for two components of work input: the one required to achieve the infinitesimal skeleton deformations $d\varepsilon_v$ and $d\varepsilon_q$ and the one associated to the infinitesimal changes in Lagrangian partial porosities $d\phi_a$ and $d\phi_w$. In this case, because the fluid phases were not connected, the internal walls of the solid skeleton delimiting the part of the porous network filled by air, as well as the internal walls delimiting the part of the porous network filled by water, remained always the same. As a result, whatever the deformation process considered the change in Lagrangian partial porosities $d\phi_a$ and $d\phi_w$ was capturing only the change in volume of the same part of the porous network filled by the respective phase (compare Fig. 1, left and right).

When fluid phases are connected, as it is the case for soils, the analysis of the contribution of each phase to the strain work $dw$ is less straightforward because the changes in Lagrangian partial porosities $\phi_w$ and $\phi_a$ do not relate only to the change of the porous volume containing the phase. Actually, in this case, $d\phi_a$ and $d\phi_w$ result from two different physical processes: a deformation process and the invasion of the volume previously containing one phase by the other phase. The invasion process is driven by the difference of pressure $u_a - u_w$ and involves finite changes of water and air contents. The deformation process is driven by both the total stress and the air and water pressures and involves infinitesimal changes of water content. The main purpose of this section is to make the distinction between these two processes as it concerns the energy input supplied to the apparent solid skeleton. It is then needed to first focus on the definition of an appropriate reference configuration related not only to the deformation, but also to the surface energy variations. The good candidate is the saturated situation where the interface energy reduces to that between the liquid and the solid particles. This is implicitly assumed in what follows.

The analysis intends to be valid for granular materials in the case sketched out in Fig. 2: the main water phase remains connected while the amount of water trapped in the intergranular menisci within the air-dominated part of the material is negligible. Accordingly the effect of these menisci is to stiffen and to strengthen the air-dominated part of the material, but the amount of water these menisci trap is not taken into account in volume balances. However we will see later on how this apparent restriction can be removed.

The reference configuration is appropriately defined by imposing a zero (atmospheric) interstitial pressure everywhere within the pore space. In the case of non connected fluids this could be achieved by a non zero air saturation as illustrated in Fig. 1-left. In the connected case this cannot be since the current water content is governed by the pressure difference $u_a - u_w$, and for a zero pressure difference the connected air and water phases cannot coexist within the porous space. Accordingly, if a drainage process is to be firstly considered the reference configuration is chosen to be fully water saturated (Fig. 2-left). In the current configuration (Fig. 2-right), the volume currently occupied by water is the pore volume delimited by the currently wetted solid grains (grey surface and black grains in Fig. 2-right). Although this volume $\phi_w V_0$ can be still expressed by (10), the current partial porosity $\phi_w$ now results from two distinct processes, namely a deformation process and a drainage process, the latter not occurring for non connected fluid phases as addressed in the previous section. In order to account separately for the contribution to $\phi_w$ related to each process let us now introduce the Lagrangian water saturation $S_r$ (Coussy, 2005; Coussy, 2006; Coussy & Monteiro, 2007; Coussy, 2007). As illustrated in Fig. 2-left the Lagrangian water saturation $S_r$ is defined in a such way that, prior to the skeleton deformation, $S_r \phi_0 V_0$ represents the volume that was delimited by the surface of the same solid grains (black grains in Fig. 2) as those that delimit the current wetted volume in the current configuration. In contrast to the Eulerian water saturation $s_r$ with a small $s$, $S_r$ with a capital $S$ is coined as the Lagrangian water saturation since it is relative to the same undeformed reference configuration. The current wetted volume $\phi_w V_0$ is finally obtained by adding to $S_r \phi_0 V_0$ the change $\varphi_w V_0$ of the porous volume resulting from the sole deformation.

$$\phi_w = s_r \phi = S_r \phi_0 + \varphi_w \qquad (18)$$

Equation (18) represents the key point of the present approach based on Lagrangian variables. It allows for splitting the current partial porosity occupied by water into the parts due to pore invasion only ($S_r \phi_0$) and deformation only ($\varphi_w$). Note that these contributions from two different physical processes cannot be decoupled by using the product $s_r \phi$ since changes in $s_r$ can be produced by both wetting/drying and deformation.

The same reasoning can be applied to the part of the porous volume filled by air, leading to the equation:

$$\phi_a = (1 - s_r)\phi = (1 - S_r) \phi_0 + \varphi_a \qquad (19)$$

where $\varphi_a$ is the change due to deformation only of the Lagrangian porosity for the part of the porous network filled by air.

It comes from Equations (18) and (19):

$$\phi = \phi_0 + \varphi_a + \varphi_w \qquad (20)$$

Equation (20) just indicates that the total volume change of the porous network sums the volume change of the parts filled by air and filled by water. It is another expression of the pure deformational essence of $\varphi_w$ and $\varphi_a$. Note that, in the case of unconnected fluid phases described in previous section, $dS_r = 0$ and thus $d\phi_w = d\varphi_w$ and $d\phi_a = d\varphi_a$. The last equation expresses again that, in absence of invasion, the changes in air and water volumetric contents are only due to deformation.

Substituting (20) into (13) the general expression of $dw$ becomes:
$$dw = d\omega + d\theta \qquad (21)$$
where
$$d\omega = p\,d\varepsilon_v + u_a\,d\varphi_a + u_w\,d\varphi_w + q\,d\varepsilon_q \qquad (22)$$
and
$$d\theta = -\phi_0\,(u_a - u_w)\,dS_r \qquad (23)$$
According to (21)-(23), the energy input $dw$ is split into two well recognizable contributions: i) a contribution $d\omega$ accounting for the skeleton strain work, that is the energy input needed to deform the solid skeleton and ii) a contribution $d\theta$ accounting for the energy input required for the invasion process to occur. For non connected fluid phases, $dS_r = 0$, which implies that $d\theta = 0$ and thus that $dw$ reduces to $d\omega$. In contrast, for a non deformable solid skeleton $d\varepsilon_v = d\varphi_a = d\varphi_w = 0$ and $dw$ reduces to $d\theta$. Let analyse now the latter case, leaving the general case of both connected fluid phases and deformable solid skeleton for the next section.

Contribution $d\theta$ is the energy input needed for displacing the air-water interface during the invasion process. When the air-water interface displaces and narrows, the variation of its free energy is caused by the removal from the interface of water molecules (de Gennes et al., 2004). As a consequence, noting $U$ the overall interface energy per unit of initial volume $V_0$, energy $d\theta$ must be equated to the infinitesimal surface energy change $dU$ due to this removal and to the replacement of the water-solid grain interfaces by the air-solid grain interfaces during the invasion process. Equality $d\theta = dU$ and equation (23) allow us to state:
$$dU = -\phi_0\,(u_a - u_w)\,dS_r \qquad (24)$$
The above relation implies that $U$ must be a function of $S_r$ only. Thus, matric suction $u_a - u_w$ is also a function of the water saturation $S_r$ only:
$$u_a - u_w = r(S_r) \qquad (25)$$
Equation (25) is the classical expression of the retention curve. This simple approach states a one-to-one relationship between the suction $u_a - u_w$ and the Lagrangian water saturation $S_r$. It does not account for hysteretic effects. These effects are generally of three origins: hydric, when the retention curve is different during a wetting process or drying process (it is generally attributed to geometrical effects such as the so-called 'ink bottle' effect); mechanical, when irreversible changes in the geometry of the porous network are caused by loading; coming from physical chemistry and then generally originating from intermolecular forces as the disjoining pressure does. Accounting for them is not contradictory to the approach presented here (see e.g. Dangla *et al.*, 1997), but requires the consideration of appropriate energy couplings that would weight down the text. They are therefore ignored in the remaining part of the paper. A brief description on how to address these effects is given in Appendix.

## THERMODYNAMICS OF PLASTICITY

*Thermodynamics bases*

Combination of the first and the second laws of thermodynamics gives the Clausius-Duhem inequality. The system considered now is the apparent solid skeleton. Since this latter is a closed system, the Clausius-Duhem inequality reads, for isothermal evolutions:
$$dD = dw - dF \geq 0 \qquad (26)$$
It expresses that, in any infinitesimal evolution, the strain work input $dw$ supplied to a system has to be greater or equal to the infinitesimal free energy $dF$ that the system can store and subsequently release in the form of useful work. The difference $dD = dw - dF$ is the dissipation spontaneously transformed into heat.

Substitution of (20) in the condition for solid incompressibility (6) provides:
$$d\varepsilon_v = -d\varphi_a - d\varphi_w \quad (27)$$
and the total strain work of the apparent solid skeleton takes the form:
$$dw = (p - u_a)\,d\varphi_a + (p - u_w)\,d\varphi_w + q\,d\varepsilon_q - \phi_0(u_a - u_w)\,dS_r \quad (28)$$
For connected phases we retrieve, as for non connected phases, that stresses $p - u_a$ and $p - u_w$ are work conjugate to the changes in air and water volumetric contents $\varphi_a$ and $\varphi_w$ due to the sole deformation process.

Using (28), the Clausius-Duhem inequality can be finally rewritten as:
$$dD = (p - u_a)\,d\varphi_a + (p - u_w)\,d\varphi_w + q\,d\varepsilon_q - \phi_0(u_a - u_w)\,dS_r - dF \geq 0 \quad (29)$$
Any further development requires the statement of dependency for the free energy $F$. It must then be recalled that $dw$ is the total strain work of the apparent solid skeleton where the contribution of the bulk air and water phases have been removed and thus that the free energy $F$ relates to a system composed by the solid skeleton and the fluid-solid interfaces only. Accordingly, and as a consequence of the additive character of energy, $F$ can be split into three parts: i) the elastic energy $\Psi$ stored in the solid skeleton during a reversible mechanical process, ii) the locked energy $Z$, which is the additional part of elastic energy that is stored in the solid skeleton when an irreversible (mechanical) process takes place and iii) the fluid-solid interfaces energy $U$ previously introduced. The concept of locked energy (also called frozen energy), early established from a formal point of view by Halphen & Nguyen (1975), has recently gained a considerable interest in soil mechanics as it allows handling the non standard character of soils within well-established thermodynamical frameworks (Coussy, 2004; Collins, 2005; Houlsby & Puzrin, 2007; Li, 2007).

For unsaturated materials, the simplest and reasonable choice for the elastic energy $\Psi$ is to assume its dependence on the elastic parts of the deformation and the Lagrangian degree of saturation $S_r$, since $S_r$ delimits the part of the solid skeleton currently subjected through the interfaces to the pressure exerted by the water phase, $u_w$. Additionally, and for the sake of simplicity, the locked energy $Z$ is assumed to depend on a unique hardening variable $\alpha$. Finally, as analysed in the previous section, the fluid-solid interfaces energy $U$ depends only on $S_r$. Denoting then by superscript p the plastic part of the deformation variables, we can write:
$$F = \Psi(\varphi_a - \varphi_a^p, \varphi_w - \varphi_w^p, \varepsilon_q - \varepsilon_q^p, S_r) + Z(S_r, \alpha) + U(S_r) \quad (30)$$
During elastic evolutions, that is when plastic deformations and hardening variable keep constant values, there is no dissipation and (29) reduces to an equality. Substitution of (30) in this equality provides the following state equations:
$$p - u_a = \frac{\partial \Psi}{\partial \varphi_a}; \quad p - u_w = \frac{\partial \Psi}{\partial \varphi_w}; \quad q = \frac{\partial \Psi}{\partial \varepsilon_q}; \quad \phi_0(u_a - u_w) = -\frac{\partial(\Psi + Z)}{\partial S_r} - \frac{dU}{dS_r} \quad (31)$$
The three first equations capture the elastic part of the behaviour of the solid matrix. The last one corresponds to the expression of the retention curve and now includes the effects of deformation, except those leading to hysteretic effects (as indicated in previous section). The first term in the right hand side of this last equation accounts for the change in free energy due to changes in water saturation at constant deformation. It is generally negligible with respect to the second term and the expression of the retention curve can be simplified into:
$$\phi_0(u_a - u_w) = -\frac{dU}{dS_r} \quad (32)$$

Equation (32) indicates that the expression of the retention curve in terms of Lagrangian degree of saturation is the same for undeformable and deformable materials (see Eq. (24)).

The state equations (31) and the dependencies considered for $F$ in Equation (30) allow for writing the Clausius-Duhem inequality (29) as:
$$dD = (p - u_a)\,d\varphi_a^p + (p - u_w)\,d\varphi_w^p + q\,d\varepsilon_q^p + \beta\,d\alpha \geq 0 \tag{33}$$
where $\beta$ is defined by:
$$\beta = -\frac{\partial Z(S_r, \alpha)}{\partial \alpha} \tag{34}$$
The variable $\beta$ is energy conjugate to the hardening variable $\alpha$ and consequently called hardening force. It will be later on associated with the current limit of elasticity. Following Equation (34), $\beta$ depends not only on the hardening variable $\alpha$, but also on the water saturation $S_r$. This distinctive point with respect to saturated conditions will appear to be crucial to model the elastoplastic response of unsaturated materials.

*Effective stress concept*

The derivation of the effective stress concept in the case of unsaturated elasticity based on thermodynamic considerations has been largely discussed in (Coussy, 2007). It will not be recalled here.

Inequality (29) indicates that, in general case, $p - u_a$ and $p - u_w$ act as independent effective stresses. To go further, additional information is needed.

Similarly to (27) the plastic incompressibility of the solid grains is now introduced. This incompressibility corresponds to plasticity due solely to irreversible sliding between undeformable solid grains and implies that:
$$d\varepsilon_v^p = -d\varphi_a^p - d\varphi_w^p \tag{35}$$
From this incompressibility relation (36), going further towards an effective stress concept requires additional assumptions. The usual assumption consists in introducing a coefficient $\chi$ ranging from 0 to 1 such as
$$d\varphi_a^p = -(1-\chi)\,d\varepsilon_v^p \; ; \; d\varphi_w^p = -\chi\,d\varepsilon_v^p \tag{36}$$
This $\chi$ factor is usually assumed equal to the degree of saturation of water $S_r$. This is a particularly questionable assumption. Indeed, as it will be seen later (Eqs. 37-38), this is to say that pores filled by water plastically deform equivalently to those filled by air whereas both groups of pores do not sustain the same pressure (the pressure difference being the capillary pressure). It is proposed here to release this restriction by assuming that the coefficient $\chi$ is no more equal to the state variable $S_r$ but only depends on $S_r$ so that $\chi = \chi(S_r)$. It should be pointed out that this is still an assumption: indeed, we may add the rate of any quantity to one of the two equations (36) and substract it from the other equation without violating (35). Nevertheless, two extreme cases do exist where the existence of the function $\chi(S_r)$ can be proved. The first case corresponds to the pore iso-deformation case, where the volumes occupied by, respectively, the water phase and the gas phase undergo the same plastic incremental deformation, so that
$$\frac{d\varphi_w^p}{\phi_0 S_r} = \frac{d\varphi_a^p}{\phi_0 (1 - S_r)} \tag{37}$$
Relations (36) and (37) then combine to give the relation
$$\chi(S_r) = S_r \tag{38}$$
This is the case grossly represented in Fig. 2, which should be seen as an illustrative example at the scale of the representative elementary volume. As said above, this is unrealistic since it would suppose that pores under air pressure experience the same plastic deformation as those under water pressure. The second case corresponds to the extreme other choice:

$$\chi(S_r < 1) = 0 \; ; \; \chi(S_r = 1) = 1 \qquad (39)$$

As defined by (39) the function $\chi(S_r)$ exhibits a non physical discontinuity of $\chi$ at $S_r = 1$. As sketched out in Fig. 3, the choice (39) must be viewed as the limit of continuously derivable functions arbitrarily chosen close to the discontinuous function defined by (39). The meaning of choice (39) may receive the following interpretation. As long as the soil remains unsaturated, that is as long as $S_r < 1$, the water phase remains discontinuous and mainly trapped within the zone delimited by the intergranular menisci. As a consequence the volume associated with the wetted zone does not evolve significantly whereas the deformation is mainly due to the deformation of the zone occupied by the air phase.

Relation (36) together with assumption $\chi = \chi(S_r)$ may be viewed as an intermediary case between the two extreme cases (38) and (39). Actually assumption $\chi = \chi(S_r)$ implies:

$$\frac{d\varphi_w^p}{\phi_0 \chi(S_r)} = \frac{d\varphi_a^p}{\phi_0 (1 - \chi(S_r))}. \qquad (40)$$

If $\phi_0 \chi(S_r) V_0$ is identified to the part of the water phase actually connected, assumption (40) stipulates that, similarly to (37), this connected part undergoes the same deformation as the corresponding apparent air phase. The non connected part of the water phase trapped within the zone delimited by the intergranular menisci, although contributing to $S_r$, has not to be accounted for in $\chi(S_r)$ so that the latter departs from $S_r$. However this non connected part affects the value of the current limit of elasticity through the relation (34) where $Z$ depends on $S_r$. Indeed, as explored later on it is at the origin of capillary hardening.

Substituting (36) into (33), we get:

$$dD = p^B d\varepsilon_v^p + q d\varepsilon_q^p + \beta d\alpha \qquad (41)$$

where $p^B$ is Bishop's stress and is defined by

$$p^B = p - [1 - \chi(S_r)] u_a - \chi(S_r) u_w \qquad (42)$$

The use of Bishop's stress is generally justified by using mixture theories or averaging procedures, starting from the microscopic momentum equations (Hassanizadeh & Gray, 1980 ; Lewis & Schrefler, 1998; Hutter *et al.*, 1999). This explains the popular, and then relevant choice $\chi = s_r$ as implied by (14). Actually the momentum equation captures the mechanical equilibrium of the current configuration where the water phase occupies the fraction $s_r$ (Eulerian water saturation) of the current porous volume. In contrast here, as for the use of Terzaghi's effective stress, the use of Bishop's stress $p^B$ in its original form (43) is justified by the incompressibility of the grains forming the solid skeleton and suggesting the definition (37) of function $\chi(S_r)$. Indeed, the sequence of relations (4), (6) to (8) leading to the concept of Terzaghi's effective stress is quite similar to the sequence of relations (26) to (28), (33), (35), (36) and (40) to (42) leading to the concept of Bishop's effective stress. In short the relation $\chi = \chi(S_r)$ as defined by (31) is a part of the plastic flow rule we are now going to explore in more details.

*Elastoplastic framework*

According to (41), Bishop's effective stress $p^B$ plays the same role in the unsaturated case as does Terzaghi's effective stress $p' = p - u$ in the saturated case. It is worth noting that this comparison is made in the sense of a unique stress thermodynamically conjugated to the deformation. Possible dependency of the hardening variable upon suction or degree of saturation is out of its scope. Furthermore, it does not burden with the possible couplings between the mechanical constitutive equations and the retention properties of a complete elastoplastic model for unsaturated soils. Such a complete model will also include a hydraulic section describing the retention properties of the material, linking the suction and the degree of saturation. The current domain of elasticity can be therefore defined by:

$$f(p^B, q, \beta) \leq 0 \tag{43}$$

where $f$ is the loading function, $q$ the deviatoric stress and $\beta$ the hardening parameter defined by Equation (34). The dependency of $\beta$ upon the degree of saturation expresses the structuring effect exerted by the intergranular menisci, which exists whatever is the connection of the water phase or the relative deformation between the pores filled by air or by water. As a consequence, $\beta$ is expressed as a function of $S_r$ and not $\chi(S_r)$.

Assuming its normality, the plastic flow rule is expressed in the form:

$$d\varepsilon_v^p = d\lambda \frac{\partial f}{\partial p^B}; \quad d\varepsilon_q^p = d\lambda \frac{\partial f}{\partial q} \tag{44}$$

where $d\lambda \geq 0$ is the plastic multiplier. The consistency condition $df = 0$ and the definition (35) of $\beta$ combine to write:

$$\frac{\partial f}{\partial p^B} dp^B + \frac{\partial f}{\partial q} dq + \frac{\partial f}{\partial \beta}\left(\frac{\partial^2 Z}{\partial \alpha^2} d\alpha + \frac{\partial^2 Z}{\partial \alpha \partial S_r^2} dS_r\right) = 0 \tag{45}$$

The hardening variable $\alpha$ varies only during plastic evolutions, that is for $d\lambda > 0$, so that $d\alpha$ has to nullify with $d\lambda$ and is finally proportional to $d\lambda$. As a consequence, consistency condition (45) allows us to express the plastic multiplier in the form:

$$d\lambda = \frac{1}{H}\left(\frac{\partial f}{\partial p^B} dp^B + \frac{\partial f}{\partial q} dq + \frac{\partial f}{\partial \beta} \frac{\partial^2 Z}{\partial \alpha \partial S_r^2} dS_r\right) \tag{46}$$

where $H$ is classical hardening modulus expressed by:

$$H = -\frac{\partial f}{\partial \beta} \frac{\partial^2 Z}{\partial \alpha^2} \frac{d\alpha}{d\lambda} \tag{47}$$

The last term on the right-hand side of Equation (46) expresses the fact that plastic strain can develop during a change in degree of saturation at constant stress. It allows modelling the phenomenon of collapse by wetting, typical of unsaturated materials.

*Modified Barcelona Model*

The elastoplastic formulation previously presented can be used to prove the thermodynamical consistency of the Barcelona Basic Model (Alonso *et al.*, 1990) in the line of the demonstration made by Coussy (2004) for the Modified Cam Clay Model.

The loading function of Modified Cam Clay Model takes the form:

$$f = \left(p' - \frac{1}{2} p_0^*\right)^2 + \frac{q^2}{M^2} - \frac{1}{4} p_0^{*2} \tag{48}$$

where $p_0^*$ is the preconsolidation pressure in saturated conditions. Its evolution with the plastic strain is given by:

$$\frac{p_0^*}{p_r} = \exp\left(\frac{1+e_0}{\lambda(0)-\kappa}\varepsilon_v^p\right) \tag{49}$$

where $e_0$ is the initial void ratio, $p_r$ the preconsolidation pressure at the reference state, $\kappa$ and $\lambda(0)$ the slope of the unloading/reloading line and saturated virgin compression line in the $e - \ln(p')$ diagram, respectively. The Barcelona Basic Model consists in extending the loading function (48) to unsaturated conditions in the form:

$$f = \left((p-u_a)+p_s-\frac{1}{2}(p_0+p_s)\right)^2 + \frac{q^2}{M^2} - \frac{1}{4}(p_0+p_s)^2 \tag{50}$$

where the preconsolidation pressure $p_0$ now also depends on the suction according to:

$$\frac{p_0}{p^c} = \left(\frac{p_0^*}{p^c}\right)^{[\lambda(0)-\kappa]/[\lambda(s)-\kappa]} \tag{51}$$

where $p^c$ is a reference pressure, $p_s$ is a tensile strength taken proportional to the current suction value $s$ and $\lambda(s)$ is the slope of the virgin compression line at suction $s$ in the $e - \ln(p-u_a)$ diagram.

Using equations (49) and (51), the preconsolidation pressure at given suction and plastic deformation may be expressed, after some rearrangements, by:

$$\frac{p_0}{p_r} = \left(\frac{p_r}{p^c}\right)^{[\lambda(0)-\lambda(s)]/[\lambda(s)-\kappa]} \exp\left(\frac{1+e_0}{\lambda(s)-\kappa}\varepsilon_v^p\right) \tag{52}$$

which can formally be re-expressed as:

$$p_0 = p_r h_s(s) h_m(\varepsilon_v^p, s) \tag{53}$$

where $h_m$ expresses the mechanical hardening due to irreversible deformations (itself affected by suction at which deformation occur) and $h_s$ represents the suction-induced hardening (which may be a function of either suction or water saturation). Use of the water retention function (26) permits to express $h_m$ and $h_s$ as functions of the (Lagrangian) degree of saturation $S_r$. Note that $h_m(\varepsilon_v^p = 0, S_r) = h_s(S_r = 1) = 1$, leading to $p_0 = p_0^* = p_r$ at a reference (saturated and non-irreversibly deformed) state.

These two models can be merged into a unique expression by using the extended Bishop's stress $p^B$ and the step function $\chi(S_r<1) = 1$ and $\chi(S_r=1) = 0$, according to

$$f = \left(p^B + p_s - \frac{1}{2}(p_0+p_s)\right)^2 + \frac{q^2}{M^2} - \frac{1}{4}(p_0+p_s)^2 \tag{54}$$

Substitution of the expression of the retention curve (25) and the hardening law (54) into (55) allows to express the preconsolidation pressure $p_0$ as a function of both $S_r$ and $\varepsilon_v^p$ and the tensile strength $p_s$ as a function of $S_r$.

Now, by identifying $p_0$ to $\beta$ and $\varepsilon_v^p$ to $-\alpha$ in the general formulation presented in the last section, the dissipation expressed by Equation (41) can be specified for the Barcelona Basic Model as:

$$dD = (p^B - p_0)\, d\varepsilon_v^p + q\, d\varepsilon_q^p \tag{55}$$

The flow rule (44) and the loading function (54) provide:

$$d\varepsilon_v^p = d\lambda \times 2\left(p^B + p_s - \frac{1}{2}(p_0+p_s)\right); \quad d\varepsilon_q^p = d\lambda \times \frac{2q}{M^2} \tag{56}$$

Substituting (56) in (55) and using the plastic loading condition, the dissipation finally reads:

$$dD = d\lambda\,(p_0+p_s)\,(p_0-p^B) \tag{57}$$

Since the plastic multiplier $d\lambda$ is always positive and $p^B$ is always lower or equal than $p_0$, the dissipation is always positive or null. As a consequence the Barcelona Basic Model is thermodynamically consistent.

Finally, it is possible to remove the discontinuity that exists in the Barcelona Basic Model as the result of the jump from $p - u_a$ to $p - u_w$ when full saturation is reached by adopting any smooth function for $\chi(S_r)$ in $p^B$ in (54). Equation (57) proves that all the models built that way are also thermodynamically consistent.

It should be noted that alternative choices for the work conjugate variables ($\alpha$, $\beta$) are possible. For instance, the modelling framework may be replaced within the theory of generalised standard materials (Lemaitre and Chaboche, 1990) or hyperplasticity (e.g. Houlsby & Puzrin, 2007). By definition, such a choice would lead to an associated evolution law for $\alpha$:

$$d\alpha = d\lambda \frac{\partial f}{\partial \beta} \qquad (58)$$

In this case, the positiveness of dissipation is automatically satisfied and does not need any particular attention. However, for the original saturated Cam-Clay model, there is still a lack of experimental evidence supporting such a choice. An interesting experimental technique that eventually may lead to some evidences about this particular concern is presented by Luong (2007) who uses infrared thermography to evaluate the energy dissipated into heat.

CONCLUDING REMARKS

In this paper, the rate equations of elastoplasticity for saturated soils have been extended to unsaturated conditions using a framework that preserved the basic laws of thermodynamics. Advance in the construction of the framework relies on the setting of several key results:

- The energy balance of the apparent solid skeleton can be split into one part due only to deformation and another part due only to pore invasion by the saturating fluids by considering as controlling variables the Lagrangian porosity (current volume of pores divided by the initial volume of porous material) and Lagrangian degree of saturation (volume, in the initial configuration, of the pores that are currently filled by water divided by the initial volume of pores).
- From the expression of the strain work, three stress variables (total stress, air pressure and water pressure) can be shown *a priori* to control the process of deformation, whereas the process of invasion is found to be controlled by the difference between air pressure and water pressure.
- The number of stress-dimension controlling variables can be reduced by assuming constraints on internal deformation. The net stress and Terzaghi's effective stress emerge as two work-conjugate variables if solid incompressibility is assumed. If pore iso-deformation is moreover assumed, the Bishop's stress (with factor $\chi$ identified to the Lagrangian degree of saturation, $S_r$) is recovered as a unique effective stress thermodynamically conjugated to the soil deformation. This last restriction can finally be relaxed by keeping $\chi$ as a smooth function of $S_r$ in the expression of Bishop's stress. This function is specific to the material under concern and relies in particular assumptions upon its microstructure.
- Elastoplastic frameworks developed for saturated soils can be extended to unsaturated conditions by setting an additional dependency of the free energy on the degree of saturation, only. In order to cope with hardening laws typical of soils in a well-posed thermodynamic framework, the free energy is split into three parts: 1) the recoverable elastic energy stored in the solid skeleton; 2) the additional locked energy stored in the solid skeleton during an irreversible mechanical process; 3) the fluid-solid interface energy. Use of Clausius-Duhem inequality allows for an unsaturated formulation where the extended Bishop's stress is the counterpart of Terzaghi's stress in the saturated formulation, even if suction (or degree of saturation) still appears in the arguments of the hardening parameter or in the definition of the water retention properties of the material.

A simple illustration of the framework is finally provided by recovering the equations of the Barcelona Basic Model. As the net stress is a limit case of the extended Bishop's stress, such a formulation allows for proving the thermodynamic consistency of the basic model as well as all the models that can be derived from it by taking a smooth function of $S_r$ as proposed in this paper. It is believed that such a framework provides the basis for further extension of more achieved models to unsaturated conditions.

APPENDIX

The water retention curve may present hysteresis effects leading to dissipation during drying-wetting cycles. In order to address these effects, let us start from Eq. (30), that is:
$$\mathrm{d}D = (p - u_a)\,\mathrm{d}\varphi_a + (p - u_w)\,\mathrm{d}\varphi_w + q\,\mathrm{d}\varepsilon_q - \phi_0\,(u_a - u_w)\,\mathrm{d}S_r - \mathrm{d}F \geq 0.$$
Substituting $F = F_{\mathrm{gr}} + U$, where $F_{\mathrm{gr}}$ stands for the free energy of the solid skeleton and $U$ is the interfaces energy, the following expression for the dissipation is obtained:
$$\mathrm{d}D = [(p - u_a)\,\mathrm{d}\varphi_a + (p - u_w)\,\mathrm{d}\varphi_w + q\,\mathrm{d}\varepsilon_q - \mathrm{d}F_{\mathrm{gr}}] + [-\phi_0\,(u_a - u_w)\,\mathrm{d}S_r - \mathrm{d}U] \geq 0.$$
The first term in brackets is the dissipation related to the solid skeleton and has been addressed in the present paper. The second term in brackets is the dissipation related to capillary hysteresis. It requires a separate treatment which is illustrated in Fig. 4 and has been addressed in Coussy (2004).

FIGURES

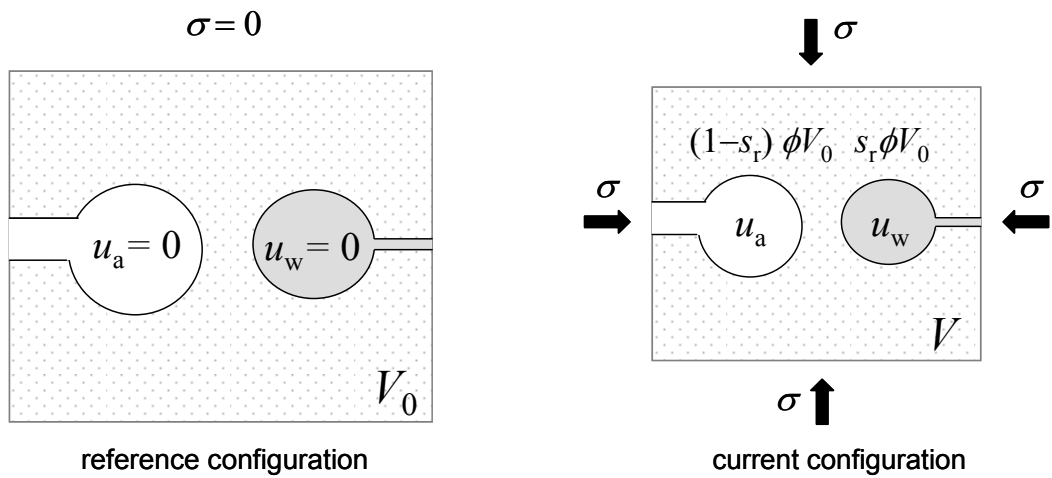

*Fig. 1. Reference configuration (left hand side) and current deformed configuration (right hand side) for non connected fluid phases.*

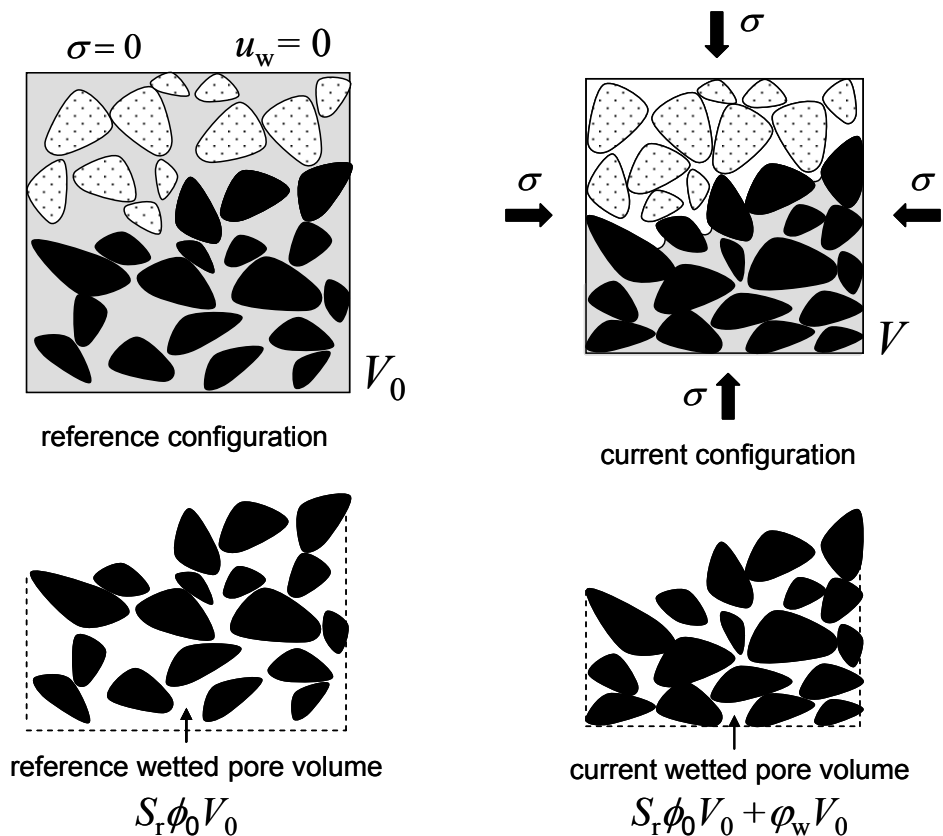

*Fig. 2. Reference configuration (left hand side) and current deformed configuration (right hand side) for connected fluid phases*

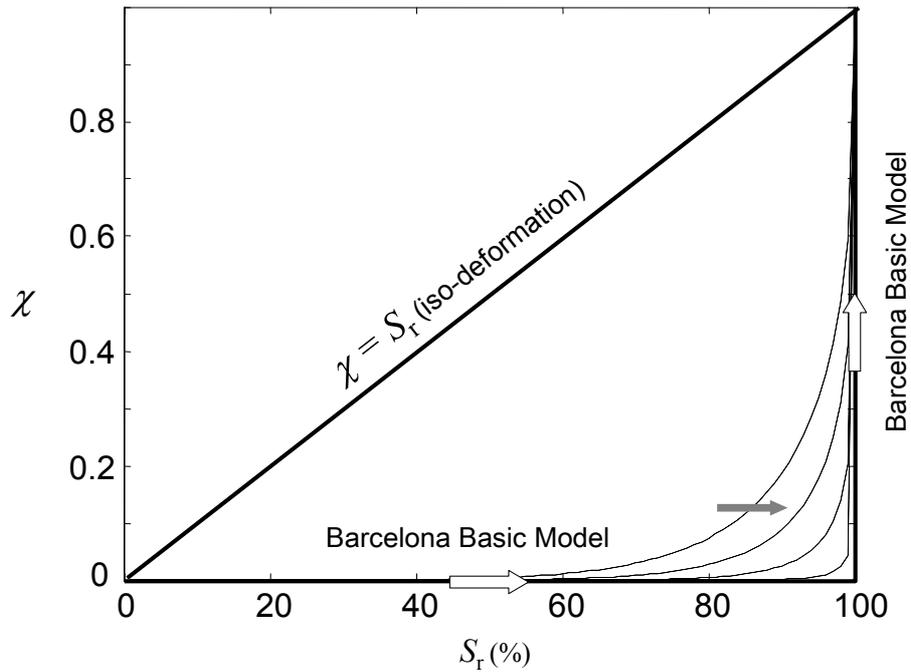

*Fig. 3. Family of functions $\chi(S_r)$ suitable for entering in the definition of an extended Bishop's effective stress. Linear function $\chi(S_r) = S_r$ define the classical Bishop's stress, step function $\chi(S_r < 1) = 0$, $\chi(1) = 1$ the pair net stress / Terzaghi's effective stress.*

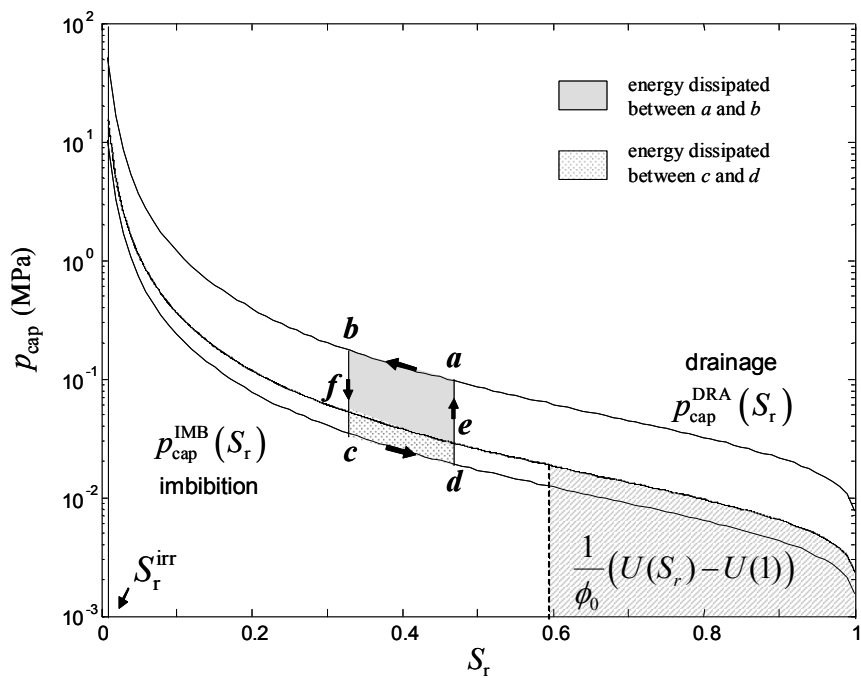

*Fig. 4. Illustration of hysteresis effects of the capillary pressure ($p_{cap}$)-degree of saturation curve and energy dissipation during a drainage-imbibition cycle.*